\documentclass[11pt,a4paper]{article}
\usepackage{jcappub}
\usepackage{color}
\usepackage{xcolor}
\usepackage{mathtools}
\usepackage{diagbox}
\usepackage{appendix}
\usepackage{amsmath,amssymb,amsfonts,dcolumn,color,graphicx,graphics,latexsym}
\usepackage{caption}
\usepackage{subcaption}
\usepackage{hyperref}
\usepackage{natbib}
\usepackage{float}
\usepackage{booktabs}

\usepackage{comment}
\usepackage{lipsum}
\usepackage{tikz}
\usepackage{multirow}
\usepackage{makecell}
\newcommand{\g}[1]{ {\color{black}  #1}}

\begin{document}

\title{Black holes in degenerate Einstein Gauss-Bonnet gravity: Can QNMs distinguish them from GR? }


 \author[a,1]{Suvikranth Gera\note{The author is currently affiliated to Department of Physics, IIT Guwahati, Guwahati, Assam, India.}}
 \emailAdd{suvikranth@iitg.ac.in}
 \affiliation[a]{Department of Physics, BITS Pilani K K Birla Goa Campus, India}
 \author[b]{Poulami Dutta Roy}
 \emailAdd{poulami@cmi.ac.in}
 \affiliation[b]{Chennai Mathematical Institute, Siruseri 603103, Tamil Nadu, India }


\abstract{ In this study, for the first time, we analyze the quasinormal modes of massless scalar fields in the context of asymptotically flat black hole solutions with zero metric determinant. These solutions were recently introduced in [JCAP $02 (2022) 02$] which satisfy degenerate Einstein Gauss-Bonnet(dEGB) action and belong to a much larger class of solutions that include cosmological constant. This solution has two distinct branches akin to Einstein Gauss-Bonnet(EBG) gravity. However, unlike the EBG solutions, both the branches of dEGB are well-defined asymptotically. The negative branch solutions from both theories are equivalent under the identification of certain metric parameters. We provide constraints on the  Gauss-Bonnet coupling parameters, which result in black hole spacetimes, and study the behaviour of a propagating scalar field through the computation of quasinormal modes. Finally, we compare the time domain evolution of the scalar field in the background of these black holes with their GR counterparts.}



\maketitle
\flushbottom

\section{ Introduction}

\noindent Over the years, there has been significant interest in formulating a gravitational framework incorporating Gauss-Bonnet(GB) and higher-order Lovelock densities in four and higher dimensions\cite{Boulware:1985wk,Castillo-Felisola:2016kpe,Deruelle:1989fj,Klein:1926tv,Madore:1985qk,Mueller-Hoissen:1985prd,Mueller-Hoissen:1985www,Wheeler:1985nh,Wheeler:1985qd}. However, these higher-order Lovelock densities with $D\leq 4$  dimensions are not dynamical. Mathematically, these terms do not contribute to the Euler-Lagrange equations of motion \cite{Lovelock:1971yv, Lovelock:1972vz,Lanczos:1938sf}. Any attempts to introduce these higher-order terms lead to additional degrees of freedom or involve derivatives higher than second order.

Recently, Glavan and Lin \cite{Glavan:2019inb} claimed that it is possible to construct a theory of gravity devoid of issues mentioned earlier while still exhibiting a dynamic influence of Gauss-Bonnet density. This framework of gravity is based on the notion of  ``singular-rescaling" of GB coupling and then taking a $D\rightarrow 4$ limit of a $D \geq 5$ action, thus resulting in an emergent theory of gravity which is four-dimensional and has non-trivial imprints of the GB coupling. However, it was later realized that this formulation is neither covariant nor the limit is well defined \cite{Gurses:2020rxb, Gurses:2020ofy, Hennigar:2020lsl, Shu:2020cjw, Ai:2020peo, Arrechea:2020evj, Arrechea:2020gjw, PhysRevD.102.024029, Mahapatra:2020rds, Hohmann:2020cor, Cao:2021nng}.  The spacetime solutions obtained by Glavan and Lin were subsequently discovered in various other frameworks of gravity \cite{Lu:2020iav, Kobayashi:2020wqy,Hennigar:2020lsl,Fernandes:2020nbq,Fernandes:2021dsb,Aoki:2020lig}.

It is imperative that any theory of gravity in four dimensions containing non-trivial effects of higher-order lovelock terms be devoid of any of the aforementioned issues. To this extent, Sengupta \cite{Sengupta:2021mpf} has proposed an emergent theory of gravity from a five-dimensional action constructed from  Einstein and Gauss-Bonnet densities with zero metric determinant. Unlike the earlier approaches, this framework does not rely on singular rescaling nor introduces any new dynamical fields. Instead, this is achieved using the notion of ``extra dimensions of zero proper length" introduced in \cite{Sengupta:2019ydf}. A similar analysis has been implemented to obtain a dynamical theory of gravity in two dimensions\cite{Gera:2021dem}. This new theory is more general than the Jackiw-Teitelboim gravity\cite{jackiw,jackiw1t} and inequivalent to the Mann-Ross prescription\cite{Mann:1992ar,mann1t1,mann1t2}. In general, metric solutions with zero determinant have been investigated as a potential resolution to curvature singularities and dark matter problems, which are serious challenges to General Relativity. Similarly, a degenerate extension to the Reissner-Nordstr\"{o}m metric has been explored in \cite{Gera:2021hei,Gera:2020fvo}, which resolves the problem of curvature singularity and provides a natural geometric interpretation of electric and magnetic charges. 
 
In this work, we study the propagation of a massless scalar wave in the background of dEGB black holes and compute the corresponding quasinormal modes (QNMs). For specific choices of parameters in the dEGB case, the scalar wave QNMs are compared with that of the GR black holes. Such a study highlights the dependence of the QNMs on the underlying theory of gravity.
Also, it is known that the solutions in higher-dimensional Einstein-Gauss-Bonnet theories in the metric formulation are plagued with instability occurring at higher multipole numbers\cite{Dotti:2005sq,Gleiser:2005ra,Konoplya:2017lhs,Konoplya:2017zwo,Takahashi:2010ye,Yoshida:2015vua,Takahashi:2011qda,Konoplya:2008ix,Takahashi:2012np,Blazquez-Salcedo:2017}.  One wonders if these solutions to degenerate EGB theory are also plagued with similar issues. Hence, this study becomes particularly pertinent as the initial step in probing any possible instabilities of the solutions in this unique dEGB theory. We also note that the 4D EGB black hole metric, whose QNMs and shadow have been studied in \cite{Konoplya:2020bxa}, matches exactly to one of the spacetimes we encounter in our work (see Sec.\ref{subsec:neg_branch}). The QNMs computed for this subclass of the solution match well with the values quoted in \cite{Konoplya:2020bxa} as shown in Appendix \ref{appendix:B} .

Lately, quasinormal modes of various compact objects have been the focus of research in multiple works where they are used to distinguish different geometries and also to identify the black hole mimicking nature of various compact objects \cite{cardoso_2019,konoplya_2016,cardoso_2016_echoes,maggio_2020,DuttaRoy:2022ytr,Roy:2021jjg,DuttaRoy:2019hij}. Such studies are especially relevant in the era of gravitational wave detections as QNMs play a crucial role in commenting on the stability of spacetime, the estimation of parameters of the remnant and also determining the underlying theory of gravity \cite{vishveshwara_1970_1,Cardoso:2019mqo,McManus:2019ulj}. On the other hand, in the context of GW data analysis, computing QNMs in alternative theories of gravity \cite{Pierini:2021jxd,Pierini:2022eim,Blazquez-Salcedo:2016enn,Blazquez-Salcedo:2017,Molina:2010fb,Pani:2013ija,Mark:2014aja,Cano:2023tmv} is crucial for formulating ringdown tests of GR \cite{Ferrari:2007dd,Berti:2009kk,Meidam:2014jpa,Brito:2018rfr,Ghosh:2021mrv,silva:2023}. These tests help constrain the possible GR violations by estimating the difference of QNMs between GR and the other theories of gravity using the ringdown signal of a binary merger event. Some works have also shown the possible detection of multi-mode ringdown signals to be a unique test of the black hole nature of the remnant formed from coalescence \cite{Dreyer:2003bv,Berti:2005ys}. The future generation detectors with their enhanced sensitivity will be an ideal ground for implementing these tests, highlighting the relevance of our study \cite{Bhagwat:2023,Berti:2016lat,Toubiana:2023}. Therefore, comparing the observational signatures, such as ringdown behaviour and QNMs of the black hole solutions in dEGB theory with that of GR is highly relevant in the era of gravitational wave detections. 

The structure of our work is as follows. In Section II, we provide a concise description of the underlying gravity theory along with its static spherically symmetric solutions, which are the focus of our study. Section III discusses the classification of these spacetimes and their properties. Section IV delves into the computation of QNMs of scalar wave propagating in dEGB spacetimes  . We conclude this paper with a summary of the results obtained and a brief perspective on the future extensions of this work in the final Section V. The main results of our work are summarized in the form of two tables; Table \ref{tab:spacetime} discusses the classification of spacetimes and Table \ref{tab:QNMsummary} shows the dependence of QNMs on metric parameters ($\kappa,Q^2$) of our theory.

\section{Revisiting the Degenerate EGB theory}

\noindent We begin by summarizing the main results of \cite{Sengupta:2021mpf}, which formulates the framework of degenerate EGB gravity. Understanding this theory of gravity is crucial before analyzing the properties and stability of its static spherically symmetric solutions. 
We begin with the action for the degenerate EGB gravity of the form
\begin{equation}
	\mathcal{L}(\hat{e},\hat{\omega})= \int dx^5 \epsilon^{\mu\nu\alpha\beta\gamma}\epsilon_{IJKLM}\left[ \frac{\alpha}{2}\hat{R}_{\mu\nu}^{IJ}(\hat{\omega})\hat{R}_{\alpha\beta}^{KL}(\hat{\omega})\hat{e}_{\gamma}^M +\frac{\zeta}{2}\hat{R}_{\mu\nu}^{IJ}(\hat{\omega})\hat{e}^{K}_{\alpha}\hat{e}^{L}_{\beta}\hat{e}_{\gamma}^{M}+\frac{\beta}{5}\hat{e}^I_\mu\hat{e}^J_\nu\hat{e}^K_\alpha\hat{e}^L_\beta\hat{e}^M_\gamma \right]
\end{equation}
where the independent fields are the vielbein $ \hat{e}^I_{\mu} $ and $ \hat{\omega}^{IJ}_{\mu} (\mu\equiv\left[ t,x,y,z,\upsilon \right], I\equiv\left[ 0,1,2,3,4 \right])$ are the super-connection. The parameters $ \alpha,\zeta,\beta $ correspond to the Gauss-Bonnet coupling, gravitational coupling, and the bare cosmological constant, respectively. Varying the action with respect to these independent fields leads to the following equations of motion
\begin{align}
\begin{split}
	&\epsilon^{\mu\nu\alpha\beta\gamma}\epsilon_{IJKLM}\left[ \alpha\hat{R}_{\mu\nu}^{IJ}(\hat{\omega})+\zeta\hat{e}^{I}_{\mu}\hat{e}_{\nu}^{J} \right]\hat{D}_\alpha(\hat{\omega})\hat{e}_{\beta}^{K}=0\\
	&\epsilon^{\mu\nu\alpha\beta\gamma}\epsilon_{IJKLM}\left[ \frac{\alpha}{2}\hat{R}_{\mu\nu}^{IJ}(\hat{\omega})\hat{R}_{\alpha\beta}^{KL}(\hat{\omega}) +\zeta \hat{R}_{\mu\nu}^{IJ}(\hat{\omega})\hat{e}^{K}_{\alpha}\hat{e}^{L}_{\beta}+ \beta \hat{e}^I_\mu\hat{e}^J_\nu\hat{e}^K_\alpha\hat{e}^L_\beta \right]=0
 \end{split}
 \label{eq:eqm}
\end{align} 
where $ \hat{D}_{\alpha}(\hat{\omega})$ represents the gauge-covariant derivative with respect to the super-connections $ \hat{\omega}_{\mu}^{IJ} $. The Bianchi identity $ \hat{D}_{\left[ \mu \right.}\hat{R}_{\left. \nu\alpha \right]}^{IJ}(\hat{\omega})=0 $ is also utilized to simplify the equations of motions and obtain them in the form stated above (eq.(\ref{eq:eqm})).

Now we restrict our analysis to the vielbeins whose determinant is zero. For this case, without  the loss of generality, we can  assume an ansatz such that the  zero eigenvalue of the vielbein $ \hat{e}^I_{\mu} $  lies along the fifth direction $ (\upsilon) $ which can be written as
\begin{align}
\begin{split}
	\hat{e}^I_{\upsilon}&=0\\
	\hat{e}^I_{\mu}&= \begin{bmatrix}
		\hat{e}^i_{a}\equiv e^{i}_{a} & 0\\
		0 & 0
	\end{bmatrix}
 \end{split}
 \label{eq:veilbein}
\end{align}
where we have used the notation of $\mu\equiv(a,\upsilon)\equiv(t,x,y,z,\upsilon)$ and $ I\equiv(i,4)\equiv(0,1,2,3,4) $.

Using the ansatz \eqref{eq:veilbein} in equations of motion \eqref{eq:eqm} leads to the following results:
\begin{itemize}
	\item An effective four-dimensional spacetime can be defined using the tetrads $ e^i_{a} $, which are invertible, unlike the five vielbein.
	\item The emergent four-dimensional spin connections satisfy the zero torsion condition. 
	\item All emergent four-dimensional fields are independent of the fifth dimension, and any such dependence is a gauge artefact.
	\item Finally, the only remaining unsolved  component of the field equations is
\begin{equation}\label{mastereq}
	\epsilon^{abcd}\epsilon_{ijkl}\left[ \phi\beta\bar{R}_{ab}^{ij}e^k_c e^l_d+\frac{\alpha}{2}\bar{R}_{ab}^{ij}\bar{R}_{cd}^{kl}+\xi e^i_a e^j_b e^k_c e^l_d \right]=0
\end{equation}
\end{itemize}
where $\bar{R}^{ij}_{ab}$ is the four dimensional torsion-free field strength tensor. Further details and analysis of these results can be found in \cite{Sengupta:2021mpf}.

\subsection{Static spherically symmetric solutions}
In order to obtain a static spherically symmetric solution to the field equation \eqref{mastereq}, we consider the ansatz
	\begin{equation}
		ds^2= -e^{\mu(r)} dt^2 + \frac{dr^2}{e^{\mu(r)}} +r^2 d\Omega^2
	\end{equation}
which can be substituted into eq.\eqref{mastereq} to obtain the  functional form of $ e^{\mu(r)} $
	\begin{equation}\label{gensolution}
		e^\mu = 1+ \frac{\phi}{2\alpha} r^2\pm \frac{1}{2}\left[\left(\frac{\phi^2}{\alpha^2}-\frac{2\chi}{\alpha}\right)r^4+ 4 C_1 r-\frac{4C_2}{\alpha}\right]^{\frac{1}{2}}
	\end{equation}
where $\alpha$ is the Gauss-Bonnet coupling and $\phi \footnote{The parameter of the theory should not be confused with the azimuthal angle $\phi$ appearing in the metric}$,$\chi$ are the theory's parameters, and $C_1$ and $C_2$ are the constants of integration. The above metric component asymptotically reduces to the form
\begin{equation}
	e^\mu\rightarrow 1-\frac{\Lambda_{eff}}{3}\pm\left(\frac{2 M_{eff}}{r}-\frac{Q_{eff}^2}{r^2}\right)
\end{equation}
where we have identified  the effective  mass, charge and  cosmological constant in terms of our metric parameters as,
\begin{align}
	& \Lambda_{eff}= \frac{-3}{2}\left(\frac{\phi}{\alpha}\pm\sqrt{\frac{\phi^2}{\alpha^2}-\frac{2\chi}{\alpha}}\right)\\
	& 2M_{eff} = \frac{C_1}{\sqrt{\frac{\phi^2}{\alpha^2}-\frac{2\chi}{\alpha}}}\\
	& Q^2_{eff} = \frac{C_2}{\alpha\sqrt{\frac{\phi^2}{\alpha^2}-\frac{2\chi}{\alpha}}}
\end{align}

Imposing the conditions that effective mass is positive and the charge is real implies $C_1<0(C_1>0)$ and $C_2<0(C_2>0)$ for the $+(-)$ branches respectively. We would also like to emphasise that $Q_{eff}$ of our theory is inequivalent to the charge originating from Maxwell Electrodynamics, as is evident from the absence of any matter fields in our action. $Q_{eff}$ is geometric in origin and in spirit similar to what has been explored in \cite{wheeler_1955,wheeler_1957}. Similar constructions of electric and magnetic charges in degenerate spacetimes have been explored in \cite{Gera:2021hei,Gera:2020fvo} where the origin of such charges was demonstrated to be associated with topological invariants. 
We will drop the subscript {\em eff} from the effective `asymptotic' parameters $\Lambda_{eff}, Q_{eff}, M_{eff}$ for brevity for the rest of the paper.

In the next section, we will focus on the asymptotically flat spacetimes in the dEGB theory and discuss their properties.




\section{Classification of solutions: with and without geometric charge}

\subsection{Solution without geometric charge: Positive branch}
Let us start our analysis by examining the positive branch of the solutions (eq.\eqref{gensolution}) given by
\begin{equation}
		e^\mu = 1+ \frac{\phi}{2\alpha} r^2+ \frac{1}{2}\left[\left(\frac{\phi^2}{\alpha^2}-\frac{2\chi}{\alpha}\right)r^4+ 4 C_1 r-\frac{4C_2}{\alpha}\right]^{\frac{1}{2}}
\end{equation}
The absence of geometric charge and asymptotic flatness would mean 
   \begin{align*}
   	& C_2=0\\
   	& \Lambda =0 \implies \frac{\chi}{\alpha}=0 ;\quad \frac{\phi}{\alpha}<0\\
   	& C_1= -2 M  \Big\vert\frac{\phi}{\alpha}\Big\vert
   \end{align*}
implementing which the metric for the positive branch can be recast as
\begin{equation}\label{eq:metric_comp}
		e^\mu= 1-\frac{r^2}{2\kappa}\left(1-\left(1-\frac{8 M \kappa}{r^3}\right)^{\frac{1}{2}}\right)
\end{equation}
with the redefinition  $\vert\frac{\alpha}{\phi}\vert=\kappa>0$. This is the form of the metric which will be used in subsequent calculations in this section.

From eq.(\ref{eq:metric_comp}), it is evident that the radial coordinate is constrained to the range $(8 M \kappa)^{\frac{1}{3}}< r<\infty$, beyond which the metric component becomes imaginary. The strictly greater limit imposed is justified by examining the exact expression of the Ricci scalar 
\begin{equation}
		R= \frac{-120 \kappa ^2 M^2+\kappa  M r^3 \left(72-48 \sqrt{1-\frac{8 \kappa  M}{r^3}}\right)+r^6 \left(6 \sqrt{1-\frac{8 \kappa  M}{r^3}}-6\right)}{\kappa  r^3 \left(r^3-8 \kappa  M\right) \sqrt{1-\frac{8 \kappa  M}{r^3}}}
\end{equation}
From the above expression, it is evident that there is a curvature singularity at $r=r_{min}=(8 M \kappa)^{\frac{1}{3}}$. To determine the position of the horizon or the infinite redshift surface for this metric we solve,
	\begin{align}
		e^{\mu}=0
		\implies -\frac{r^2}{2k}\left(1-\frac{8M\kappa} {r^3}\right)^{\frac{1}{2}}=1-\frac{r^2}{2\kappa}\label{eq:+vebranchmasshorizon}
	\end{align}
Squaring both sides of eq.\eqref{eq:+vebranchmasshorizon} reduces it to a quadratic polynomial equation of the form
\begin{equation}
	r^2-2M r -\kappa=0
\end{equation}
having solutions
\begin{equation}
	r_\pm = M\pm\sqrt{M^2+\kappa}
\end{equation}
Only the $r_+$ is admissible of the two roots as $r_-$ leads to negative radial values for $\kappa>0$.
However, note that the above solutions correspond to the squared equation, which may or may not satisfy the original equation in eq.\eqref{eq:+vebranchmasshorizon}. 

Let us further analyse the eq.\eqref{eq:+vebranchmasshorizon} at the horizon $r=r_h$,
\begin{equation}
		-\frac{r_h^2}{2k}\left(1-\frac{8M\kappa}{r_h^3}\right)^{\frac{1}{2}}=1-\frac{r_h^2}{2\kappa}
\end{equation}
Since  $\kappa>0$ and the square root term gives only a positive value, the LHS is always negative. Hence the above equation is only satisfied if 
\begin{equation}
	1-\frac{r_h^2}{2\kappa}= \frac{\kappa-2M^2-2M\sqrt{M^2+\kappa}}{2\kappa}\leq 0
\end{equation}
with $r_h$ expressed in terms of $\kappa$ and giving
\begin{equation}
	\kappa-2M^2-2M\sqrt{M^2+\kappa}\leq 0
\end{equation}
The above inequality is satisfied for $\kappa\leq 8M^2$, where the equality is achived at  $\kappa = 8M^2$. Beyond this range of $\kappa$ the inequality is violated and  there are no horizons. Note that  for $ \kappa = 8M^2 $, the event horizon coincides with the curvature singularity, forming a naked singularity. We will focus on cases with $\kappa < 8 M^2$ to only study the black hole solutions.

\subsection{Solution without geometric charge: Negative  branch}
We now focus on the solution which has the negative sign in eq.\eqref{gensolution}, corresponding to the negative branch having the form
\begin{equation}
	e^\mu = 1+ \frac{\phi}{2\alpha} r^2- \frac{1}{2}\left[\left(\frac{\phi^2}{\alpha^2}-\frac{2\chi}{\alpha}\right)r^4+ 4 C_1 r-\frac{4C_2}{\alpha}\right]^{\frac{1}{2}}
\end{equation}
 As with the previous case, we impose the  conditions of asymptotic flatness and  zero geometric charge which, for the negative branch, are given by
\begin{align*}
	C_2=&\,0\\
	\Lambda =&0 \implies \frac{\chi}{\alpha}=0; \quad \frac{\phi}{\alpha}>0\\
	C_1=& \, 2 M  \frac{\phi}{\alpha}
\end{align*}
Implementing the above conditions along with the parameter $\kappa=\frac{\alpha}{\phi}>0$ in the negative branch, the  metric function becomes
\begin{equation}
	e^\mu= 1+\frac{r^2}{2\kappa}\left(1-\left(1+\frac{8 M \kappa}{r^3}\right)^{\frac{1}{2}}\right)
\end{equation}
We will use the above form of the  metric function for the rest of the calculations in this section. Note that this solution matches the metric explored in \cite{Glavan:2019inb}, with the identification of $\kappa$ to the GB coupling in \cite{Glavan:2019inb}. However, the range of $\kappa$ does not match the range of GB coupling in \cite{Glavan:2019inb} due to the conditions on the constraints being unrelated. We believe this correspondence is coincidental, as these solutions do not match in presence of a cosmological constant. For completeness, we proceed with our  analysis of the  metric properties and in the later sections will explore their corresponding QNMs associated with scalar wave propagation.

The Ricci scalar for this case has a form
\begin{equation}
	R=\frac{120 \kappa ^2 M^2+\kappa  M r^3 \left(72-48 \sqrt{\frac{8 \kappa  M}{r^3}+1}\right)+r^6 \left(6-6 \sqrt{\frac{8 \kappa  M}{r^3}+1}\right)}{\kappa  r^3 \left(8 \kappa  M+r^3\right) \sqrt{\frac{8 \kappa  M}{r^3}+1}}
\end{equation}
which, unlike the positive branch, is singular only at $r=0$. Hence the range of radial coordinate is unobstructed, and $r$ takes the values from $0<r<\infty$.
After finding the position of singularity, we move on to check the presence of horizons by solving
\begin{align}\label{eq:-vebranchmainmasshorizon}
	&e^\mu=0 \implies 1+\frac{r^2}{2\kappa}=\frac{r^2}{2\kappa}\left(1+\frac{8M\kappa}{r^3}\right)^{\frac{1}{2}}
\end{align}
Squaring both sides gives the horizon at
\begin{equation}\label{eq:-vebranchmasshorizon}
	r_\pm=M\pm\sqrt{M^2-\kappa}
\end{equation}
Since $\sqrt{M^2-\kappa}<M$, we have two horizons corresponding to the two solutions. $\kappa= M^2$ results in an extremal case where both the horizons merge, producing a black hole solution with a single horizon. Hence for the negative branch, the allowed range of parameter $\kappa$ is $0<\kappa\leq M^2$. Note that unlike the positive branch for the range $r>0$, the LHS and RHS of \eqref{eq:-vebranchmainmasshorizon} are always positive; hence all the values of $\kappa$ in the range $0<\kappa\leq M^2$ are valid solutions.

\subsection{Charged static spherically symmetric solution: Positive branch}
We will now deal with the solutions having a non-zero geometric charge $Q$. Starting with the positive branch, we have the metric of the form
	
\begin{equation}
	e^\mu = 1+ \frac{\phi}{2\alpha} r^2+ \frac{1}{2}\left[\left(\frac{\phi^2}{\alpha^2}-\frac{2\chi}{\alpha}\right)r^4+ 4 C_1 r-\frac{4C_2}{\alpha}\right]^{\frac{1}{2}}
\end{equation}
Similar to the cases without geometric charge, we impose the asymptotic flatness giving rise to conditions
\begin{align*}
		& \Lambda =0 \implies \frac{\chi}{\alpha}=0 ;\quad \frac{\phi}{\alpha}<0\\
			& \frac{C_2}{\alpha}=- Q^{2}\vert\frac{\phi}{\alpha}\Big\vert \\
		& C_1= -2 M  \Big\vert\frac{\phi}{\alpha}\Big\vert
\end{align*}
which are used to recast the metric as,
\begin{equation}
	e^\mu= 1-\frac{r^2}{2\kappa}\left(1-\left(1-\frac{8 M \kappa}{r^3}+\frac{4Q^{2}\kappa}{r^{4}}\right)^{\frac{1}{2}}\right)
\end{equation}
with $\vert\frac{\alpha}{\phi}\vert=\kappa>0$. This is the metric form that will be used in subsequent calculations involving the positive branch with non-zero $Q$. 

The Ricci scalar  for the above metric can be written as
	\begin{equation}
		R= \frac{F(r,M,Q,\kappa)}{\kappa  r^8 \left(\frac{-8 \kappa  M r+4 \kappa  Q^2+r^4}{r^4}\right)^{3/2}}
	\end{equation}
where the exact form of $F(r,M,Q)$ is not relevant for us since we focus only on the position of curvature singularity given by the roots of  $ r^4-8 \kappa  M r+ 4 \kappa  Q^2=0$. This quartic equation has to be solved numerically for each value of $(\kappa, M, Q)$ giving the position of curvature singularity and hence the allowed range of the radial coordinate.

Similar to the previous cases, the position of the horizon is determined by solving
	\begin{align}
		e^{\mu}=0
		\implies -\frac{r^2}{2k}\left(1-\frac{8M\kappa} {r^3}+\frac{4Q^{2}\kappa}{r^{4}}\right)^{\frac{1}{2}}=1-\frac{r^2}{2\kappa}\label{eq:+vebranchchargehorizon}
	\end{align}
which on squaring reduces to a quadratic equation
	\begin{equation}
		r^2-2M r -\kappa+ Q^{2}=0
	\end{equation}
having solutions
	\begin{equation}
		r_\pm = M\pm\sqrt{M^2-Q^{2}+\kappa}
  \label{eq:+vechargehorizon}
	\end{equation}
Depending on the value of $ \kappa $ and $ Q^{2} $, the metric can have one or two positive roots and hence horizons, but in the context of QNMs (which is the focus of the current work), we are only interested in the outer horizon located at $r=r_+$. Note that if we demand the presence of a horizon in spacetime, we get a constraint on the geometric charge
\begin{equation}\label{eq:+vebranchchargeconstraint}
		Q^{2}\leq M^{2}+\kappa
\end{equation}
Following the analysis done in the case without geometric charge, we determine if all the solutions given in eq.(\ref{eq:+vechargehorizon}) actually satisfy the horizon condition of eq.\eqref{eq:+vebranchchargehorizon}.

\begin{figure}[h]
\centering
\includegraphics[scale=1]{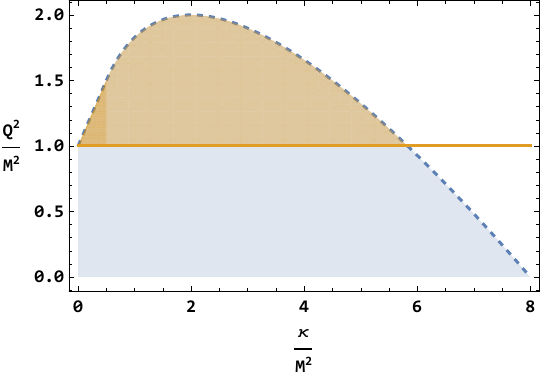}
\caption{Shaded region represents allowed parameter range of $\kappa$ and $Q^{2}$ for black hole solutions. The yellow region depicts extreme charge to mass ratios, while the parameters on the dotted lines lead to naked singularities.}
\label{fig:+vebranchchargeparameterrange}
\end{figure}

At the horizon $r=r_h$, eq.\eqref{eq:+vebranchchargehorizon} becomes 
	\begin{equation}
		-\frac{r_h^2}{2k}\left(1-\frac{8M\kappa}{r_h^3}+\frac{4 \kappa Q^{2}
	 }{r^{4}}\right)^{\frac{1}{2}}=1-\frac{r_h^2}{2\kappa}
	\end{equation}
The LHS of the above equation is always negative since  $\kappa>0$ and the term under square root gives only a positive value. Hence the equation is satisfied only if 
\begin{equation}
		1-\frac{r_h^2}{2\kappa}= \frac{\kappa-2M^2+Q^{2}-2M\sqrt{M^2-Q^{2}+\kappa}}{2\kappa}\leq 0
\end{equation}
with $r_h$ expressed in terms of $\kappa$, leading to the condition
\begin{equation}
		\kappa-2M^2+Q^2-2M\sqrt{M^2-Q^2+\kappa}\leq 0
  \label{eq:Q_dependence_1}
\end{equation}
The above inequality is satisfied for $0<\kappa\leq\frac{1}{2}M^2$ provided the charge is constrained as suggested in eq.\eqref{eq:+vebranchchargeconstraint}. However, for $\kappa\geq\frac{1}{2}M^2$, we get additional condition on charge given by 
	\begin{equation}
		Q^{2}\leq 2\sqrt{2\kappa M^2}-\kappa\leq M^2+\kappa
  \label{eq:Q_dependence_2}
	\end{equation}
In Fig.\ref{fig:+vebranchchargeparameterrange}, we show the allowed range of metric parameters in the parameter space of $(Q^2,\kappa)$. The shaded region corresponds to black hole solutions while the parameter values falling on the dotted line give naked singularity. The yellow shaded region denotes black holes with $Q^2/M^2 \geq1$ where the equality corresponds to the extremal case. Note that  $Q$ beyond the extremal value is allowed in this degenerate theory of gravity.
	
\subsection{Charged static spherically symmetric solution: Negative Branch}
Once again, we analyse the negative branch of the solution in the presence of a non-zero geometric charge having the metric
\begin{equation}
		e^\mu = 1+ \frac{\phi}{2\alpha} r^2- \frac{1}{2}\left[\left(\frac{\phi^2}{\alpha^2}-\frac{2\chi}{\alpha}\right)r^4+ 4 C_1 r-\frac{4C_2}{\alpha}\right]^{\frac{1}{2}}
\end{equation}
To ensure asymptotic flatness of the spacetime, the following conditions must be imposed on the parameters
	\begin{align}
		& \Lambda=0 \implies \frac{\chi}{\alpha}=0 ;\quad \frac{\phi}{\alpha}>0\\
			& \frac{C_2}{\alpha}=Q^{2}\frac{\phi}{\alpha} \\
		& C_1= 2 M  \frac{\phi}{\alpha}
	\end{align}
giving the metric as
	\begin{equation}
		e^\mu= 1+\frac{r^2}{2\kappa}\left(1-\left(1+\frac{8 M \kappa}{r^3}-\frac{4Q^{2}\kappa}{r^{4}}\right)^{\frac{1}{2}}\right)
	\end{equation}
with $\frac{\alpha}{\phi}=\kappa>0$. The Ricci scalar for this case is given by
	\begin{equation}
		R= \frac{F(r,M,Q,\kappa)}{\kappa  r^8 \left(\frac{8 \kappa  M r-4 \kappa  Q^2+r^4}{r^4}\right)^{3/2}}
	\end{equation}
where, similar to the positive branch case, we do not concern with the exact form of $F(r,M,Q)$. We calculate the roots of $ r^4 +8 \kappa  M r - 4 \kappa  Q^2=0$ numerically and find the position of curvature singularity which gives the allowed range of radial coordinate for specific choice of $(
\kappa,Q^2)$.
To find the horizon, we solve	
	\begin{align}
		e^{\mu}=0
		\implies \frac{r^2}{2k}\left(1+\frac{8M\kappa} {r^3}-\frac{4Q^{2}\kappa}{r^{4}}\right)^{\frac{1}{2}}=1+\frac{r^2}{2\kappa}
  \label{eq:-vechargetrans}
	\end{align}
which on squaring reduces to a quadratic equation
	\begin{equation}
		r^2-2M r + \kappa+ Q^{2}=0
	\end{equation}
with solution
	\begin{equation}
		r_\pm = M\pm\sqrt{M^2-Q^{2}-\kappa}
  \label{eq:-vechargehorizon}
	\end{equation}
Depending on the value of $( \kappa , Q^{2} )$, the metric can have one or two positive roots and hence horizons, but we will focus only on the outer horizon $r_+$ since we will deal with purely ingoing wave at horizon as part of the boundary conditions for QNMs. We also obtain the following constraint on the geometric charge,
	\begin{equation}\label{eq:-vebranchchargeconstraint}
		Q^{2}\leq M^{2}-\kappa
	\end{equation}
which ensures presence of horizon in the spacetime.	
To check whether $r_{\pm}$ of eq.(\ref{eq:-vechargehorizon}) are indeed the solutions to transcendental equations of eq.\eqref{eq:-vechargetrans}, we evaluate them at the horizon $r=r_h$
\begin{equation}
		\frac{r_h^2}{2k}\left(1+\frac{8M\kappa}{r_h^3}-\frac{4 \kappa Q^{2}
	 }{r^{4}}\right)^{\frac{1}{2}}=1+\frac{r_h^2}{2\kappa}
\end{equation}
	Since  $\kappa>0$ and the square root term gives only a positive value, the LHS is always positive. Hence the above equation is only satisfied if 
	\begin{equation}
		1-\frac{r_h^2}{2\kappa}= \frac{2M^2-Q^2+ 2M\sqrt{M^2-Q^2-\kappa}}{2\kappa}\geq 0
	\end{equation}
where we have expressed $r_h$ in terms of $\kappa$ which reduces to
	\begin{equation}
		2M^2-Q^2+ 2M\sqrt{M^2-Q^2-\kappa}\geq 0
	\end{equation}
The above inequality is automatically satisfied for the range $0<\kappa\leq M^2$ and $Q^2\leq M^2-\kappa$, hence any value $\kappa$ and $Q$ within the parameter range is allowed.\\
 
Table \ref{tab:spacetime} summarizes the properties and parameter ranges for the black hole solutions discussed in this section corresponding to positive and negative branch. In the next segment of this paper, we will study the propagation of scalar wave in the black hole spacetimes and the associated QNMs.

\begin{table}[h]
\footnotesize
  \hspace*{-1.5cm}
  \begin{tabular}{|c|c|c|c|c|}
      \hline
      & Branch  & Metric & Horizon & Allowed parameters \\ [0.5 cm]\hline 
  \multirow[c]{2}{3em}{Q = 0} & $+$   & $e^\mu= 1-\frac{r^2}{2\kappa}\left(1-\left(1-\frac{8 M \kappa}{r^3}\right)^{\frac{1}{2}}\right)$  &  $r_\pm = M\pm\sqrt{M^2+\kappa}$ &  $0\leq \kappa \leq 8 M^2$\\ [0.5cm] \cline{2-5}
                     & $-$   & $e^\mu= 1+\frac{r^2}{2\kappa}\left(1-\left(1+\frac{8 M \kappa}{r^3}\right)^{\frac{1}{2}}\right)$  & $r_\pm=M\pm\sqrt{M^2-\kappa}$ & $0<\kappa\leq M^2$\\ [0.5cm]
                     \hline
                     \hline
     \multirow[c]{2}{3em}{Q $\neq$ 0} & $+$   & $e^\mu= 1-\frac{r^2}{2\kappa}\left(1-\left(1-\frac{8 M \kappa}{r^3}+\frac{4Q^{2}\kappa}{r^{4}}\right)^{\frac{1}{2}}\right)$  & $r_\pm = M\pm\sqrt{M^2-Q^{2}+\kappa}$ & \makecell{$Q^{2}\leq 2\sqrt{2\kappa M^2}-\kappa,\quad 0<\kappa\leq \frac{M^2}{2}$ \\  $Q^2\leq M^2+\kappa,\quad \frac{M^2}{2}<\kappa< 8 M^2$} \\ [0.5cm] \cline{2-5}
                     & $-$   & $e^\mu= 1+\frac{r^2}{2\kappa}\left(1-\left(1+\frac{8 M \kappa}{r^3}-\frac{4Q^{2}\kappa}{r^{4}}\right)^{\frac{1}{2}}\right)$  & $r_\pm = M\pm\sqrt{M^2-Q^{2}-\kappa}$ & $0<\kappa\leq M^2$, $Q^2\leq M^2-\kappa$\\ [0.5cm]
                     \hline                 
\end{tabular}
\caption{\label{tab:spacetime}Table shows asymptotically flat black hole solutions in the dEGB theory, their horizon position, and allowed parameter ranges.}
\end{table}

\section{Stability analysis: Scalar wave propagation and quasinormal modes}

In this section, we will probe the behavior of a propagating massless scalar wave and the corresponding QNMs in the background of our black hole spacetimes studied in the previous section. Quasi-normal modes are complex characteristic frequencies associated with a perturbed object, through which it radiates energy to reach a stable state and are accompanied with purely outgoing boundary conditions at infinity. Each QNM has a structure, $\omega = \omega_r +i \omega_i$, where the real part denotes the oscillation frequency and the imaginary part corresponds to the inverse of the damping time. In our convention, the damped signal is defined by  $|\omega_i|<0$. Physically, this represents a stable scalar wave. A more robust probe of the stability of the black hole spacetimes is through gravitational perturbations.  The existence of a particular spacetime solution necessitates its stability under tensor perturbations. However, in the context of the degenerate framework of gravity, the perturbations of the vielbein fields are not well understood and require dedicated attention, which is beyond the scope of this paper. Hence, we restrict our analysis to the propagation of massless scalar fields in black hole spacetime backgrounds. 

We begin with the Klein-Gordon equation for a massless scalar field,
 \begin{align}
    \Box \Phi = 0.
    \label{eq:KG}
\end{align}

Note that the scalar field $\Phi$ is a test field that does not directly perturb the background metric. However, the propagation and the behaviour of the scalar field is influenced by the background geometry. As the background spacetime is spherically symmetric and static,  we decompose $\Phi$ in terms of spherical harmonics, where the indices of $Y(\theta, \phi)$ have been suppressed for simplicity,
\begin{align}
  \Phi(t,r,\theta,\phi)=Y(\theta,\phi)\frac{u(r)e^{-i \omega t}}{r}.
\end{align} 

Incorporating this ansatz in eq.(\ref{eq:KG}), we obtain the radial equation of the form of a Schr\"{o}dinger-like equation by using the tortoise coordinate $r_*\left( dr_*=\frac{dr}{e^{\mu}}\right)$,
\begin{align}
  \frac{d^2 u}{d r_*^2}+[\omega^2 - V_{eff}]u=0
  \label{eq:radial_eqn}
 \end{align}
where the effective potential is
\begin{align}
    V_{eff}(r)= e^\mu \Big(\frac{\ell(\ell+1)}{r^2}+\frac{\left(e^\mu\right)'}{r} \Big).
    \label{eq:potential_r}
\end{align}
The derivative is with respect to the radial coordinate and $\ell$ is the azimuthal number arising from the separation of variables. In the forthcoming sections, we will plot the effective potential for specific solutions and parameter values and comment on their nature. To compute the discrete quasinormal modes, we implement the well-studied WKB technique along with Pad{\'e} improvements \cite{iyer_1987, konoplya_wkb_2003,konoplya_wkb_2019} and also plot the time evolution of the massless scalar field in the spacetime background. 

The semi-analytical scheme of computing the QNMs of a single barrier potential using the WKB approximation was developed by Schutz and Will \cite{schutz_1985}. The WKB method with first order in series, gives a simple analytical formula relating the QNM frequencies to the effective potential,
\begin{equation} \label{eq:WKB}
    \omega^2 = V_0 - i(p+\frac{1}{2}) \sqrt{-2 V_0^{''}}, \,\, p=0,1,2...
\end{equation}
where $V_0$ and $V_0^{''}$ denote the value of the potential and its second derivative, with respect to the radial coordinate, at the point of maxima of the potential, respectively. The overtone number is denoted by `p' with $p=0$ being the fundamental mode. In our work, we will restrict to the most dominant fundamental mode for a particular angular momentum. Through multiple works in literature, the WKB technique has been extended to $13^{th}$ order along with the inclusion of Pad{\'e} approximants to improve accuracy (see \cite{konoplya_wkb_2003, konoplya_wkb_2019, konoplya_2011} and the references therein). Depending on the nature of the spacetime being studied, the most stable WKB order is chosen, which is not necessarily the highest order in the series expansion. For our spacetimes, we apply the $6^{th}$ order WKB with Pad{\'e} approximants $(\Tilde{m} = 5)$. The WKB method is suitable for higher angular momentum modes,\g{such that $p << \ell$ where $p$ is the overtone number. For our study, we focus on fundamental modes only with $p=0$, and hence implement the WKB to cases with higher values of $\ell$}. The stability of the scalar field needs to be verified for small angular momentum modes such as  $\ell=0$ and $\ell=1$,\g{for which the WKB technique is less accurate}. To this extent, we plot the time domain evolution of the scalar field. The damping behaviour of these time domain profiles would indicate the stability of the propagating scalar field.

In order to study the time evolution of the scalar field $\Phi(t,r,\theta,\phi) = Y(\theta,\phi)\psi(r,t)$, we take the radial part of the scalar wave equation and keeping the derivative in time we recast the equation for $\psi(r,t)$ into the following form\cite{PhysRevD.49.883,konoplya_2011,DuttaRoy:2019hij},
\begin{align}
    \frac{\partial^{2}\psi}{\partial t^{2}} - \frac{\partial^{2}\psi}{\partial r_*^{2}} + V_{eff}(r_*) \psi =0 .
    \label{eq:QNM}
\end{align}
The above equation can be written using the light cone coordinates $(u,v)$ such that $du=dt-dr_*, dv= dt+dr_*$ and, 
\begin{equation}
	\Big(4 \frac{\partial^{2}}{\partial u \partial v} + V(u, v) \Big) \psi(u,v) = 0.
	\label{eq:uv_equation}
\end{equation}

Following \cite{konoplya_2011}, we define the operator of time evolution in these coordinates as,
\begin{equation}
	\begin{split}
		exp\Big(h \frac{\partial}{\partial t}\Big) = exp\Big( h\frac{\partial}{\partial u} + h \frac{\partial}{\partial v} \Big) 
		= exp \Big(h \frac{\partial}{\partial u}\Big) + exp\Big( h\frac{\partial}{\partial v}\Big) -1\\ + \frac{h^{2}}{2} \Big( exp\Big( h\frac{\partial}{\partial u}\Big) + exp\Big(h \frac{\partial}{\partial v} \Big) \Big) \frac{\partial^{2}}{\partial u \partial v} + ....
	\end{split}    
\end{equation}
which when applied on $\psi (u,v)$ and taking into account eq.(\ref{eq:uv_equation}) gives, 
\begin{equation}
	\psi(u+h,v+h) = \psi(u+h,v) + \psi(u, v+h) - \psi(u,v) - \frac{h^{2}}{8} V(u,v) (\psi(u+h,v)+ \psi(u, v+h))
	\label{eq:discretization}
\end{equation}
where $h$ denotes the step size of the discretization scheme. A grid is constructed on two null-surfaces $u=u_{0}$ and $v=v_{0}$ and using 
eq.(\ref{eq:discretization}) the value of $\psi$ is calculated at each grid point starting from the initial data specified on $(u_0,v_0)$. The initial condition $\psi (u,0)= e^{\frac{-(u-6)^2}{100}}$ and $\psi(0,v) =$ constant, the value of which is determined by $\psi(0,0)$, is defined on the null grid. The Gaussian pulse is evolved along the grid and plotted at a constant value of the spatial coordinate as a function of time. The time domain profile, thus obtained, is a superposition of the QNMs of the scalar field, dampening of which would indicate its stability. The profile should be invariant under the change of grid size $h$, thus being free from any numerical artefacts. Further details about the discretization scheme and the relation between $(u,v)$ and $(t,r_*)$ can be found in \cite{konoplya_2011}.

The semi-analytical technique for computing QNMs is complimented with a plethora of numerical tools which are discussed in multiple works in literature (see \cite{konoplya_2011, berti_2009} and the references therein for an overview of different techniques). One such numerical technique is the Prony extraction \cite{konoplya_2011,DuttaRoy:2019hij,Roy:2021jjg} in which  the time domain data is fitted with damped sinusiods. We implement this technique to verify the QNMs obtained from the WKB method and is discussed in Appendix \ref{appendix:prony}. In our QNM analysis, we set $M=1$ unless otherwise stated.

\subsection{QNMs of positive branch solution without geometric charge}
\begin{figure}[h]
\centering
\begin{subfigure}{0.49\textwidth}
	\includegraphics[width=\linewidth]{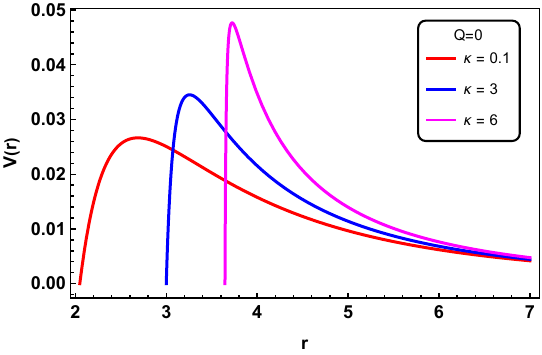}
     \caption{V(r) for positive branch, $\ell=0$}
     \label{fig:zero_Q_pos_branch_l0}
 \end{subfigure}
 \begin{subfigure}{0.49\textwidth}
 	\includegraphics[width=\linewidth]{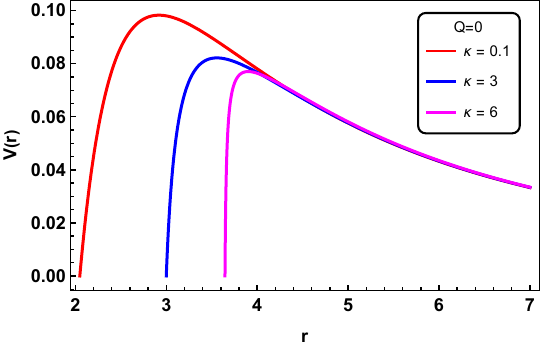}
     \caption{V(r) for positive branch, $\ell=1$}
     \label{fig:zero_Q_pos_branch_l1}
 	\end{subfigure}
  \par\bigskip
 \begin{subfigure}{0.49\textwidth}
	\includegraphics[width=\linewidth]{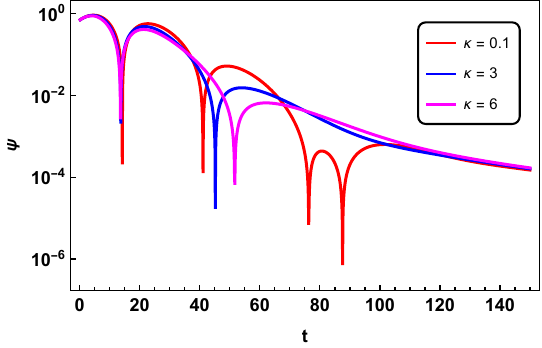}
      \caption{TD profile for $\ell=0$}
      \label{fig:zero_Q_pos_branch_l0_TD}
 \end{subfigure}
 \begin{subfigure}{0.49\textwidth}
 	\includegraphics[width=\linewidth]{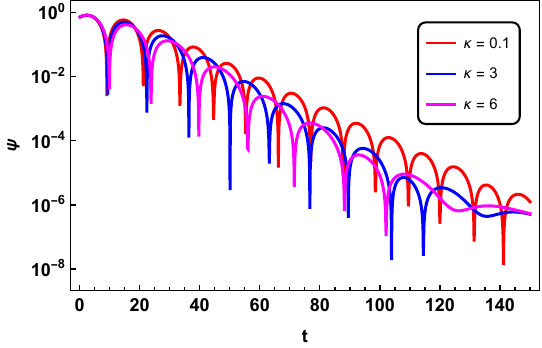}
     \caption{TD profile for $\ell=1$}
      \label{fig:zero_Q_pos_branch_l1_TD}
 	\end{subfigure}
 \caption{Top panel: The variation of effective potential V(r) with the radial coordinate for the positive branch spacetimes with different values of $\kappa$ and ${\rm Q=0}$. The angular momentum mode is $\ell=0$ and 1. A single barrier potential is observed for all the spacetimes considered. Bottom panel: Time domain (TD) profile for the above mentioned spacetimes with different $\kappa$ as observed at $r_{*}=6, \ell=0,1$. The initial Gaussian profile $\psi= e^{\frac{-(u-6)^2}{100}}$ is evolved over time with grid spacing of 0.1. The damped TD profile indicates the stability of the scalar field. }
 \end{figure}

\begin{figure}[h]
\centering
\begin{subfigure}{0.6\textwidth}
 \centering
	\includegraphics[width=\linewidth]{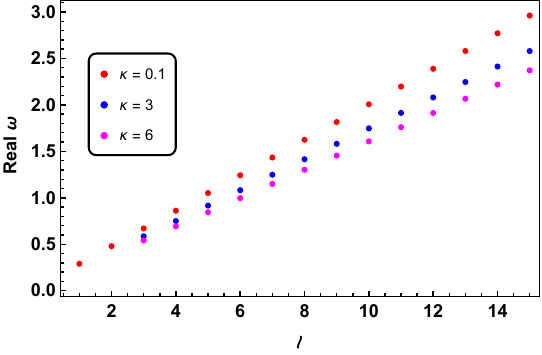}
    \caption{Variation of Re($\omega$) with $\ell$}
  \label{fig:pos_branch_zero_Q_realQNM}
 \end{subfigure}
 \par\bigskip
\begin{subfigure}{0.49\textwidth}
	\includegraphics[width=\linewidth]{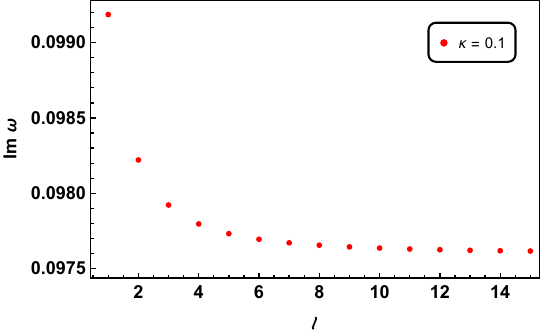}
  \caption{Variation of Im($\omega$) with $\ell$, $\kappa=0.1$}
    \label{fig:pos_branch_zero_Q_imQNM1}
 \end{subfigure}
 \begin{subfigure}{0.49\textwidth}
 	\includegraphics[width=\linewidth]{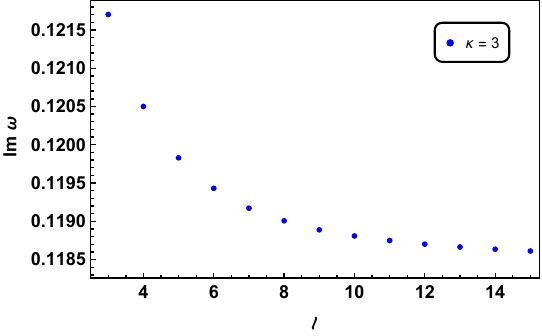}
   \caption{Variation of Im($\omega$) with $\ell$, $\kappa=3$}
      \label{fig:pos_branch_zero_Q_imQNM2}
 	\end{subfigure}
 	\caption{The variation of the real (upper panel) and imaginary (bottom panel) part of the fundamental QNM with the angular momentum mode is shown for the positive branch case with zero geometric charge. The QNMs are calculated using the WKB method till $6^{th}$ order with Pad\'e approximation ($\Tilde{m} =5)$ for different values of $\kappa $. The magnitude of Re($\omega$) (Im ($\omega$)) decreases (increases) with increasing $\kappa$ for a fixed $\ell$. }
 \end{figure} 
We begin by examining the structure of the effective potential obtained in eq.(\ref{eq:potential_r}) for various values of $\kappa$ belonging to the positive branch case and plot it as a function of the radial coordinate as 
shown in Fig.(\ref{fig:zero_Q_pos_branch_l0}) and (\ref{fig:zero_Q_pos_branch_l1}). It is evident from the plots that the potential is a single barrier for  $\kappa <8$. Hence, the WKB method can be applied to determine the QNMs. The time domain profiles, for different $\kappa$, with $\ell =0$ shown in Fig.(\ref{fig:zero_Q_pos_branch_l0_TD}) and $\ell =1$ shown in Fig.(\ref{fig:zero_Q_pos_branch_l1_TD}), indicate the stability of the scalar field due to their decaying nature over time. To understand the behavior of the fundamental QNMs and their dependence on the metric parameter $\kappa$, we plot the variation of the real part of the modes for different $\ell$ and $\kappa$ in Fig.(\ref{fig:pos_branch_zero_Q_realQNM}), and the imaginary part in Fig.(\ref{fig:pos_branch_zero_Q_imQNM1}) and (\ref{fig:pos_branch_zero_Q_imQNM2}). A detailed study of the above mentioned figures lead to the following observations:

\begin{itemize}
\item We observe from Fig.(\ref{fig:pos_branch_zero_Q_realQNM}) that as the angular momentum mode increases, the real component of the QNMs also increases, indicating a rise in frequency. Although distinct, the magnitudes of the modes are very close for different values of $\kappa$ especially for small $\ell$.  Also, as mentioned earlier, the WKB method \g{works well} for higher values of $\ell$. Hence, we have not computed the QNMs corresponding to very low $\ell$ values for $\kappa = 3,5,6$. In order to verify the stability of the scalar field for $\ell=0,1$, we plot the time domain profiles as shown in Fig.(\ref{fig:zero_Q_pos_branch_l0_TD}) and (\ref{fig:zero_Q_pos_branch_l1_TD}). The damped nature indicates a stable scalar field evolution even for low $\ell$ values.
\item The imaginary component, on the other hand, decreases with $\ell$ as shown in Fig.(\ref{fig:pos_branch_zero_Q_imQNM1}) and (\ref{fig:pos_branch_zero_Q_imQNM2}) for all values of $\kappa$. This indicates that the higher modes have longer damping time and can dominate the signal at later stages.
\item For a fixed angular momentum mode, the frequency and damping time decreases as $\kappa$ increases. This is evident from the corresponding time domain profiles for different $\kappa$ where the time domain signal for $\kappa=6$ decays much faster than smaller $\kappa$ geometries.
\end{itemize}  
\begin{table}[h]
  \centering
  \begin{tabular}{|c|c|c|c|c|}
      \hline
     $\ell$ &  Schwarzschild & $\kappa=0.001$ &  $\kappa=0.01$ & $\kappa=0.1$\\ 
    \hline 
  1 & 0.292909 -i 0.09776 & 0.292928 -i 0.097405 & 0.292692 -i 0.097818 & 0.290542 -i 0.099187\\
  2 & 0.48364 -i 0.096757 & 0.483610 -i 0.096733 & 0.483274 -i 0.096908 & 0.480022 -i 0.0982177\\
  3 & 0.675365 -i 0.096499 & 0.675436 -i 0.098474 & 0.674859 -i 0.096646 & 0.670409 -i 0.097922\\
  \hline
\end{tabular}
\caption{\label{tab:+vebranchmasscompare} Comparison of fundamental QNM for different $\kappa$ values corresponding to the positive branch of $Q=0$ case with that of the Schwarzschild black hole. The QNMs are calculated using the WKB method till $6^{th}$ order with Pad\'e approximation ($\Tilde{m} =5)$. For smaller $\kappa$, the spacetimes have comparable QNMs to that of the Schwarzschild black hole.}
\end{table}

For a vanishing coupling constant, the dEGB theory reduces to GR with the only vacuum, static, spherically symmetric solution being the Schwarzschild case.
We thus compare the scalar QNMs generated by a Schwarzschild black hole with that of our solution with a small coupling constant. This analysis will help us visualize the difficulty of segregating the two kinds of solutions only through their scalar QNMs and highlight the QNMs' dependence on the underlying gravity theory. Table \ref{tab:+vebranchmasscompare} shows that the fundamental QNMs for small $\kappa$ are significantly close (upto the accuracy considered here) to the corresponding QNMs of Schwarzschild black hole for a particular angular momentum mode. Hence, in actual observation, a highly sensitive detector will be required to distinguish the solution of our theory with a small $\kappa$ from a Schwarzschild black hole.

\begin{figure}[h]
 \centering
    \begin{subfigure}{0.49\textwidth}
 \centering
	\includegraphics[width=\linewidth]{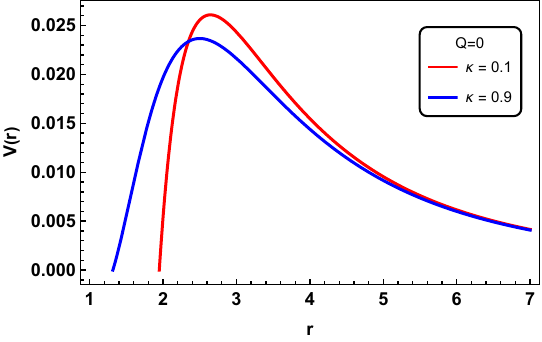}
     \caption{V(r) for negative branch, $\ell=0$}
     \label{fig:zero_Q_neg_branch_l0}
 \end{subfigure}
 \begin{subfigure}{0.49\textwidth}
 	\includegraphics[width=\linewidth]{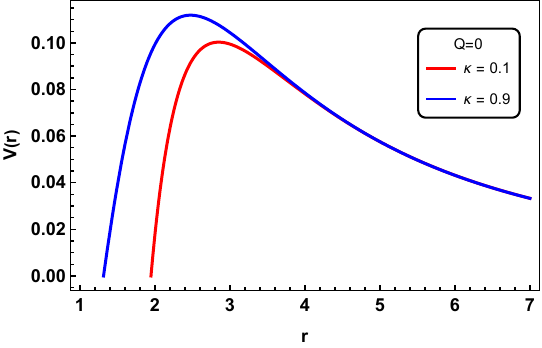}
     \caption{V(r) for negative branch, $\ell=1$}
     \label{fig:zero_Q_neg_branch_l1}
 	\end{subfigure}
  \par\bigskip
  \begin{subfigure}{0.49\textwidth}
 \centering
	\includegraphics[width=\linewidth]{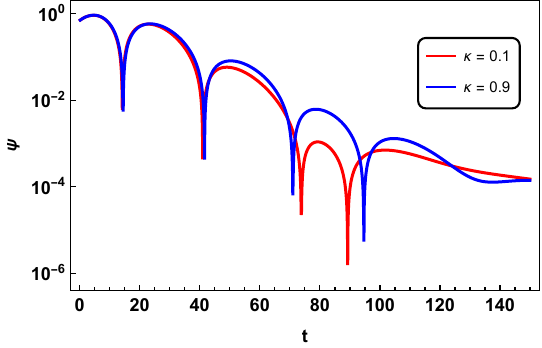}
      \caption{TD profile for $\ell=0$}
      \label{fig:zero_Q_neg_branch_l0_TD}
 \end{subfigure}
 \begin{subfigure}{0.49\textwidth}
 	\includegraphics[width=\linewidth]{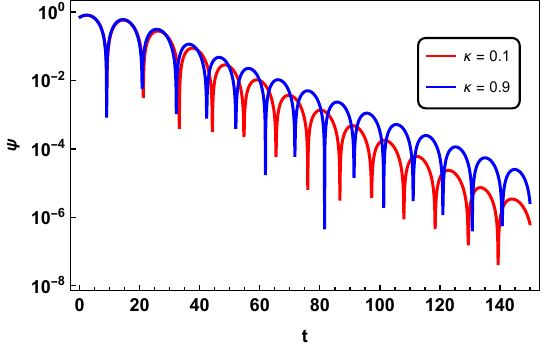}
    \caption{TD profile for $\ell=1$}
     \label{fig:zero_Q_neg_branch_l1_TD}
 	\end{subfigure}
 	\caption{Top panel: The variation of effective potential V(r) with the radial coordinate for the negative branch, $Q=0$ spacetimes. The angular momentum mode is $\ell=0$ and 1. A single barrier potential is observed for all the spacetimes considered. Bottom panel: Time domain (TD) profile for the above mentioned spacetimes with different $\kappa$  as observed at $r_{*}=6, \ell=0,1$. The initial Gaussian profile $\psi= e^{\frac{-(u-6)^2}{100}}$ is evolved over time with grid spacing of 0.1. The damped TD profile indicates stability of the scalar field. }
 \end{figure}
\subsection{QNMs of negative branch solution without geometric charge} 
\label{subsec:neg_branch}

\begin{figure}[h]
\centering
\begin{subfigure}{0.6\textwidth}
	\includegraphics[width=\linewidth]{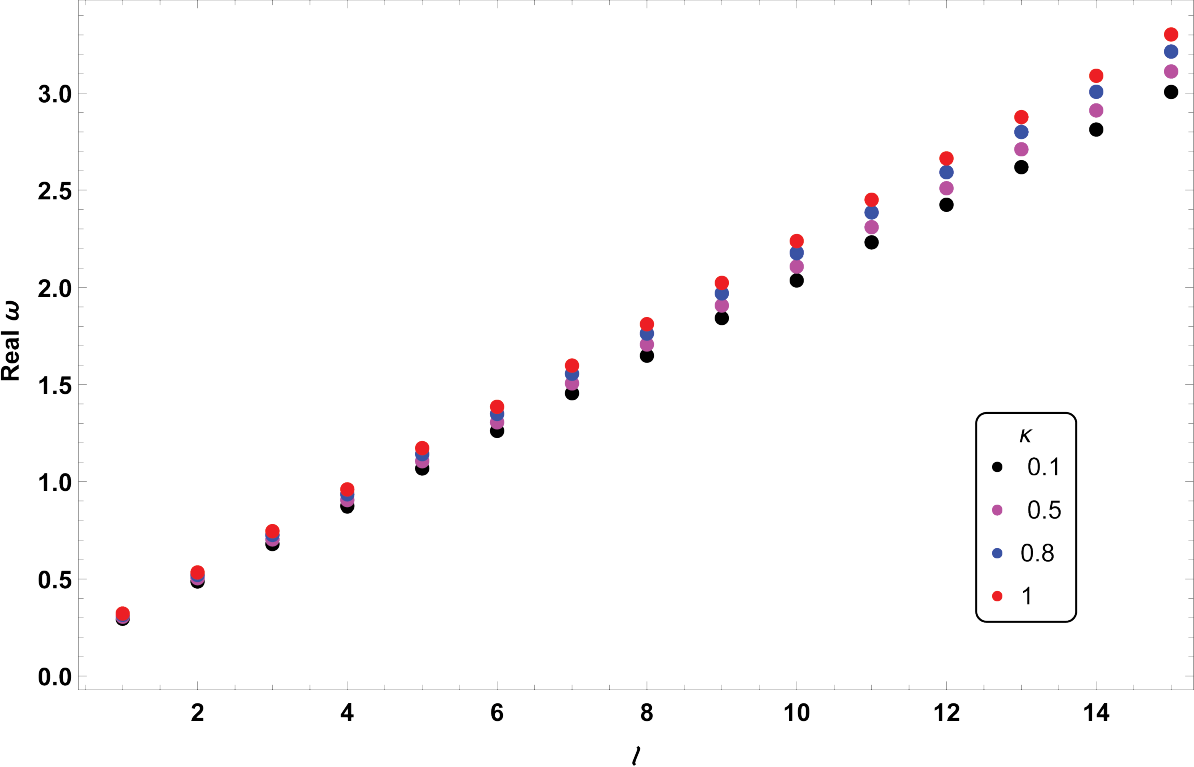}
 	\caption{Variation of Re($\omega$) with $\ell$}
 	\label{fig:neg_branch_zer_Q_realQNM}
 \end{subfigure}
 \par\bigskip
\begin{subfigure}{0.49\textwidth}
	\includegraphics[width=\linewidth]{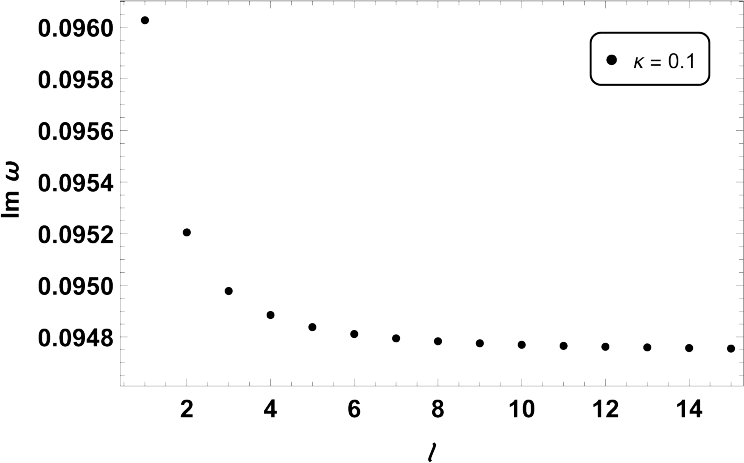}
 	\caption{Variation of Im($\omega$) with $\ell$, $\kappa =0.1$}
 	\label{fig:neg_branch_zer_Q_imQNM_1}
 \end{subfigure}
 \begin{subfigure}{0.49\textwidth}
 	\centering
 	\includegraphics[width=\linewidth]{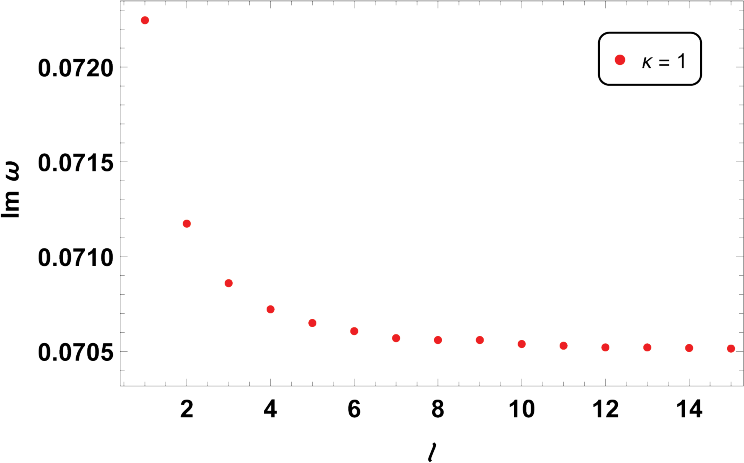}
 	\caption{Variation of Im($\omega$) with $\ell$, $\kappa =1$}
 	\label{fig:neg_branch_zer_Q_imQNM_2}
 	\end{subfigure}
 	\caption{The variation of the real (upper panel) and imaginary (bottom panel) part of the fundamental QNM with the angular momentum mode is shown for the negative branch case with zero geometric charge. The QNMs are calculated using the WKB method till $6^{th}$ order with Pad\'e approximation ($\Tilde{m} =5)$ for different values of $\kappa $. The magnitude of Re($\omega$) (Im ($\omega$)) increases (decreases) with increasing $\kappa$ for a fixed $\ell$.}
 \end{figure}
Similar to the positive branch solution, we will compute the scalar quasinormal modes associated with the negative branch case by solving eq.(\ref{eq:radial_eqn}). A single barrier potential is observed for different $\kappa$ as shown in Fig.(\ref{fig:zero_Q_neg_branch_l0}) and (\ref{fig:zero_Q_neg_branch_l1}). Thus the WKB method with Pad\'e improvements can be applied to compute the fundamental QNMs. To verify the stability of the scalar field for small angular momentum modes, we plot the time domain profile with $\ell=0$ and $\ell=1$ as shown in Fig.(\ref{fig:zero_Q_neg_branch_l0_TD}) and (\ref{fig:zero_Q_neg_branch_l1_TD}), respectively. A damped profile for both cases indicates stability. 

We move on with the study of the QNMs and their dependence on metric parameters. Fig.(\ref{fig:neg_branch_zer_Q_realQNM}) shows the variation of the real part of the QNMs as a function of $\ell$ for different $\kappa$. In contrast to the positive branch case, the real component of the fundamental mode increases with increasing $\kappa$. The imaginary part, on the other hand, decreases as $\kappa$ increases, as shown in Figs.(\ref{fig:neg_branch_zer_Q_realQNM}) and (\ref{fig:neg_branch_zer_Q_imQNM_1}). This behavior of QNMs is also evident from the time domain profiles as shown in Fig(\ref{fig:zero_Q_neg_branch_l0_TD}) and (\ref{fig:zero_Q_neg_branch_l1_TD}). Note that the metric of the negative branch with zero geometric charge is equivalent to the metric proposed in \cite{Glavan:2019inb} under the identification of $\alpha$ with parameter $2\kappa$ of our case. The scalar QNMs and shadows of this metric have been studied in \cite{Konoplya:2020bxa}. We verify our results for the scalar field QNMs with the ones quoted in \cite{Konoplya:2020bxa} and find excellent agreement (see Appendix \ref{appendix:B}).

As done for the positive branch, we compute the QNMs for different $\kappa$ values and compare them with the Schwarzschild case. Table \ref{tab:-vebranchmasscomp} shows that small $\kappa$ solutions indeed have almost identical QNMs to that of the Schwarzschild black hole, with the distinction becoming more prominent as $\kappa$ increases.
\begin{table}[h]
  \centering
  \footnotesize
  \begin{tabular}{|c|c|c|c|c|c|}
      \hline
     $\ell$ &  Schwarzschild & $\kappa=0.001$ &  $\kappa=0.1$ & $\kappa=0.7$ & $\kappa = 0.9$\\ 
    \hline 
  1 & 0.292909 -i 0.09776 & 0.2931761 -i 0.0975015 & 0.295398 -i 0.096027& 0.3122758 -i 0.083077 & 0.318868 -i 0.076427 \\
  2 & 0.48364 -i 0.096757 & 0.484015 -i 0.096608 & 0.487438 -i 0.0952057 & 0.51529 -i 0.082522 &  0.527432 -i 0.075695\\
  3 & 0.675365 -i 0.096499 & 0.6758755 -i 0.096352 & 0.680578 -i 0.094978 & 0.719381 -i 0.082369 & 0.736788 -i 0.075478 \\
  \hline
\end{tabular}
\caption{\label{tab:-vebranchmasscomp} Comparison of fundamental QNM for different $\kappa$ values belonging to the negative branch with zero geometric charge to that of the Schwarzschild black hole. The QNMs are calculated using the WKB method till $6^{th}$ order with Pad\'e approximation ($\Tilde{m} =5)$. For smaller $\kappa$, the spacetimes have comparable QNMs to that of the Schwarzschild black hole.}
\end{table}

\subsection{QNMs of positive branch solution with geometric charge}
\begin{figure}[h!]
\begin{subfigure}{0.49\textwidth}
	\includegraphics[width=\linewidth]{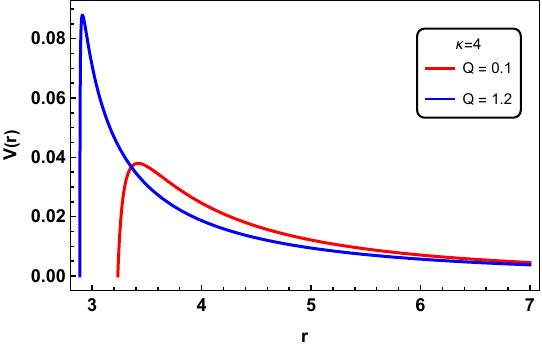}
 	\caption{V(r) for $\ell =0$}
  \label{fig:pos_branch_Q_pot_l0}
 \end{subfigure}
 \begin{subfigure}{0.49\textwidth}
 	\includegraphics[width=\linewidth]{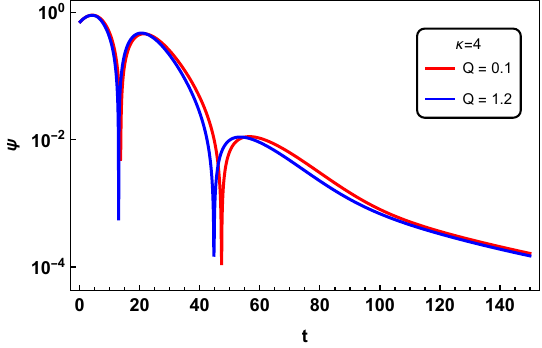}
 	\caption{TD profile for $\ell =0$}
   \label{fig:TD_pos_branch_Q_l0}
 	\end{subfigure}
 	\caption{Left panel: The variation of effective potential V(r) with the radial coordinate for the positive branch spacetimes with $\kappa=4$ and different values of $Q$. The angular momentum mode is $\ell=0$. A single barrier potential is observed for all the spacetimes considered. Right panel:Time domain (TD) profile for $\kappa = 4$ corresponding to the above mentioned spacetimes as observed at $r_{*}=6,\ell=0,1$ with initial Gaussian profile $\psi= e^{\frac{-(u-6)^2}{100}}$ and grid spacing 0.1. The damped TD profile indicates the stability of the scalar field.}
 \end{figure}

We study the potentials corresponding to various parameter values for a representative $\ell$ mode as shown in Fig.(\ref{fig:pos_branch_Q_pot_l0}). The effective potentials are found to be a single barrier for all $(\kappa,Q)$ hence QNM frequencies can be computed via WKB technique. For brevity, we show the plot for one arbitrary choice of parameter values, but the qualitative single barrier structure is present for all $\ell, \kappa$ and $Q$. 
The WKB techniques are known to be suitable for larger values of $\ell$ and may not provide convergent results for lower values as discussed in the previous sections. Thus, the stability of the scalar wave, while propagating in the dEGB black hole spacetime with non-zero geometric charge, is established via time-domain profiles for small $\ell$. For example, we demonstrate the decaying of the signal for $\kappa=4, \ell=0$ in Fig.(\ref{fig:TD_pos_branch_Q_l0}). Note that the choice of $\kappa$ and $Q$ to demonstrate this is arbitrary, and the objective result regarding the stability of the scalar field for this branch is independent of the choice of parameters in the allowed range, i.e. the scalar field is stable for the lower $\ell$ values for the complete range of allowed metric parameters. 

 \begin{figure}[h]
 \begin{subfigure}{0.49\textwidth}
	\includegraphics[width=\linewidth]{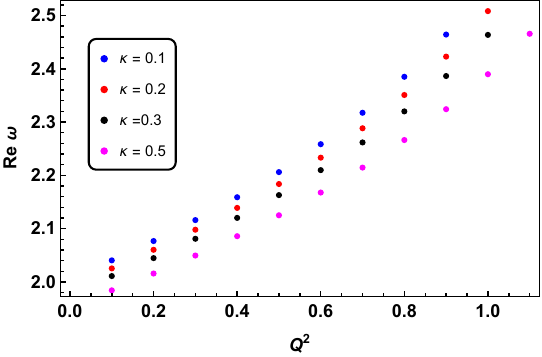}
    \caption{Variation of Re($\omega$) with $Q^2$ for small $\kappa$ }
    \label{fig:pos_branch_Q_small_kappa_realQNM}
 \end{subfigure}
 \begin{subfigure}{0.49\textwidth}
	\includegraphics[width=\linewidth]{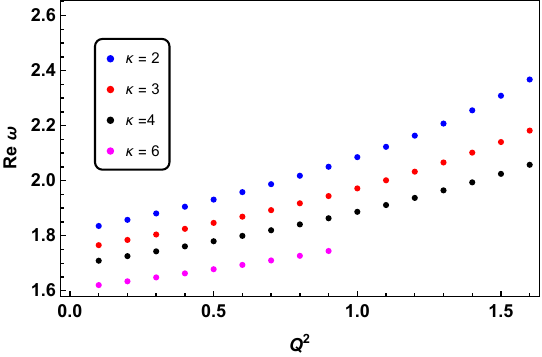}
    \caption{Variation of Re($\omega$) with $Q^2$ for large $\kappa$ }
    \label{fig:pos_branch_Q_large_kappa_realQNM}
 \end{subfigure}
 \par\bigskip
 \begin{subfigure}{0.49\textwidth}
	\includegraphics[width=\linewidth]{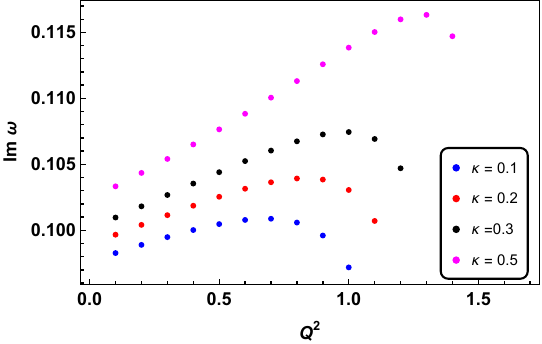}
    \caption{Variation of Im($\omega$) with $Q^2$ for small $\kappa$ }
    \label{fig:pos_branch_Q_small_kappa_imQNM}
 \end{subfigure}
 \begin{subfigure}{0.49\textwidth}
	\includegraphics[width=\linewidth]{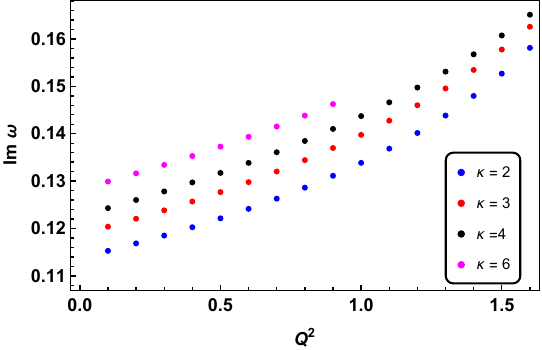}
    \caption{Variation of Im($\omega$) with $Q^2$ for large $\kappa$ }
    \label{fig:pos_branch_Q_large_kappa_imQNM}
 \end{subfigure}
 \caption{The variation of the real (upper panel) and imaginary (bottom panel) part of the fundamental QNM with $Q^2$ is shown for the positive branch case and $\ell=10$. The QNMs are calculated using the WKB method till $6^{th}$ order with Pad\'e approximation ($\Tilde{m} =5)$ for different values of $(\kappa,Q)$. The magnitude of Re($\omega$) increases with increasing $Q^2$ for both small and large values of $\kappa$. The Im($\omega$) increases then decreases with $Q^2$ for small $\kappa$ spacetimes but increases monotonically for large $\kappa$.}\label{fig:+vebranchchargeQNM}
 \end{figure}

Since the QNMs now depend on both $Q$ and $\kappa$, we plot the fundamental modes for a fixed $\ell$ value, $\ell=10$, and study their dependence on the metric parameters as shown in Fig.(\ref{fig:+vebranchchargeQNM}). Certain observations can be made about the QNMs for the positive branch solutions.
\begin{itemize}
\item In Fig.(\ref{fig:pos_branch_Q_small_kappa_realQNM}) and (\ref{fig:pos_branch_Q_large_kappa_realQNM}), we observe the dependence of real part of QNM on geometric charge $Q$, for small and large values of $\kappa$, respectively. For a fixed $\kappa$, the frequency increases as the charge increases. On the other hand, as $\kappa$ increases, the frequency decreases for a fixed $Q$. Note that the allowed values of $Q$ changes according to the $\kappa$ considered, following eq.(\ref{eq:Q_dependence_1}) and (\ref{eq:Q_dependence_2}). 
\item  However, the imaginary component has some interesting properties as shown in Fig.(\ref{fig:pos_branch_Q_small_kappa_imQNM}) and (\ref{fig:pos_branch_Q_large_kappa_imQNM}) for small and large $\kappa$, respectively. For small $\kappa$, the plot of imaginary $\omega$ vs $Q^2$ has the same characteristic behaviour as the Reissner-Nordstr\"{o}m black hole where the magnitude of Im($\omega$) increases, reaches a peak and then decreases with $Q^2$. Thus for one particular $\kappa$, the shortest damping time will correspond to  specific $Q^2$ lying in the middle of the allowed charge range. However, this changes for higher values of $\kappa$ for which the imaginary component increases with an increase in charge, indicating faster signal damping for large $Q^2$. 
\end{itemize}

Similar to the case of zero geometric charge, we now compare the fundamental QNMs of the positive branch black holes, for different values of $\kappa$, with that of the Reissner-Nordstr\"{o}m (RN) black hole, as shown in Table \ref{tab:chargeQNMcompare}. As expected, smaller $\kappa$ corresponding to a weaker coupling term in dEGB theory gives QNM values closer to the RN case.

\begin{table}[h]
\footnotesize
  \centering
  \begin{tabular}{|c|c|c|c|c|c|c|}
      \hline
                   &       &     &   \multicolumn{2}{c|}{Positive branch}           &   \multicolumn{2}{c|}{Negative branch}              \\ \cline{4-7}
     $\ell$  &Q  &Reissner-Nordstr\"{o}m & $\kappa=0.01$ & $\kappa=0.1$ & $\kappa=0.01$ & $\kappa=0.1$ \\ \hline
  \multirow[c]{2}{2em}{1}  & 0.1     &  0.2934 -i 0.09781  & 0.2931 -i 0.09795 &  0.2911 -i 0.09922   & 0.2936 -i 0.09766 &  0.2957 -i 0.09635 \\ \cline{2-7}
                           & 0.9     &  0.3526 -i 0.09721 &  0.352 -i 0.09726 & 0.347 -i 0.10171 &    0.3531 -i 0.09671&   0.3584 -i 0.09141\\ \hline
                     
  \multirow[c]{2}{2em}{2}  & 0.1     & 0.4844 -i 0.09681   & 0.4841 -i 0.09696 & 0.4808 -i 0.09828  & 0.4848 -i 0.09666 & 0.4883 -i 0.09524 \\ \cline{2-7}
                           & 0.9     & 0.5819 -i 0.09663 & 0.5809 -i 0.09711 &  0.5726 -i 0.10096               & 0.5829 -i 0.09614 &  0.5924 -i 0.09109 \\ \hline
                     
  \multirow[c]{2}{2em}{3}  & 0.1     & 0.6765 -i 0.09655 & 0.6759 -i 0.0967 & 0.6715 -i 0.09798 & 0.677 -i 0.09641 & 0.6817 -i 0.09502   \\ \cline{2-7}
                           & 0.9     & 0.8125 -i 0.09647 & 0.8111 -i 0.09694 &  0.7995 -i 0.10074          & 0.8139 -i 0.09598 &  0.8275 -i 0.09096 \\ 
  \hline
\end{tabular}
\caption{\label{tab:chargeQNMcompare} Comparison of fundamental QNM for the positive and negative branch spacetimes with different values of $\kappa$ and $Q$ to that of the Reissner-Nordstr\"{o}m black hole. The QNMs are calculated using the WKB method till $6^{th}$ order with Pad\'e approximation ($\Tilde{m} =5)$. For smaller $\kappa$, the spacetimes have comparable QNMs to that of the RN black hole.}
\end{table}

\subsection{QNMs of negative branch solution with geometric charge}
\begin{figure}[h]
\begin{subfigure}{0.45\textwidth}
	\includegraphics[width=\linewidth]{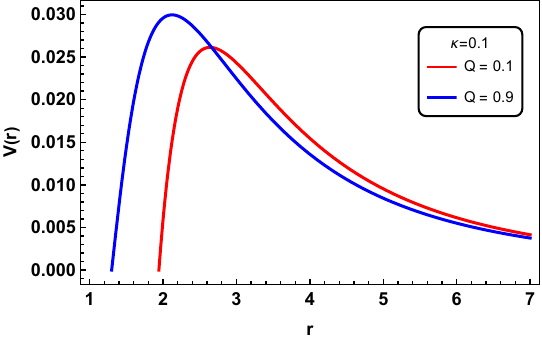}
 	\caption{V(r) for $\ell =0$}
  \label{fig:neg_branch_Q_pot_l0}
 \end{subfigure}
 \begin{subfigure}{0.45\textwidth}
 	\includegraphics[width=\linewidth]{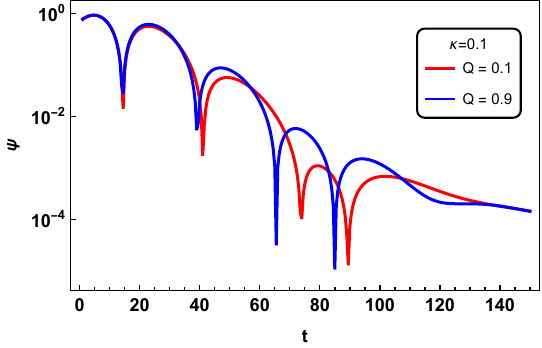}
 	\caption{TD profile for $\ell =0$}
   \label{fig:TD_neg_branch_Q_l0}
 	\end{subfigure}
 	\caption{Left panel: The variation of effective potential V(r) with the radial coordinate for the negative branch spacetimes with $\kappa=0.1$ and different values of $Q$. The angular momentum mode is $\ell=0$. A single barrier potential is observed for all the spacetimes considered. Right panel: Time domain (TD) profile corresponding to the above mentioned spacetimes as observed at $r_{*}=6,\ell=0,1$ with initial Gaussian profile $\psi= e^{\frac{-(u-6)^2}{100}}$ and grid spacing 0.1. The damped time domain profile indicates stability of the scalar field and the associated background geometries.}
 \end{figure}

As with the other previous solutions, we move on to discuss the  potentials for various values of $Q$ and Gauss-Bonnet coupling and compute the corresponding QNMs under the propagation of the scalar field. The effective potential is found to be single barrier for all $\ell, \kappa$ and $Q$. For an arbitrary choice of parameters, we show the plot of the effective potential as a function of radial coordinate in Fig.(\ref{fig:neg_branch_Q_pot_l0}). The corresponding time domain profile for $\ell=0$ shown in Fig.(\ref{fig:TD_neg_branch_Q_l0}) indicates stability of the scalar field for small angular momentum modes. 
\begin{figure}[h]
	\begin{subfigure}{0.48\textwidth}
		\includegraphics[width=\linewidth]{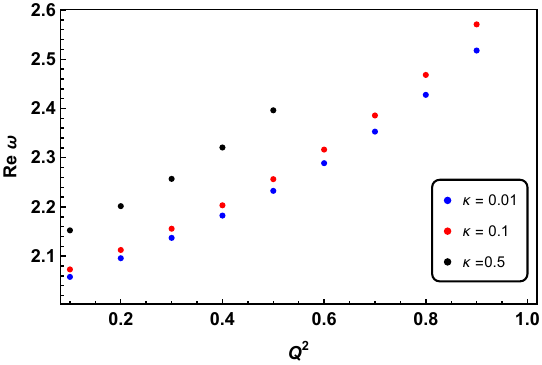}
  \caption{Variation of Re($\omega$) with $Q^2$}
  \label{fig:neg_branch_Q_realQNM}
	\end{subfigure}
	\begin{subfigure}{0.50\textwidth}
		\includegraphics[width=\linewidth]{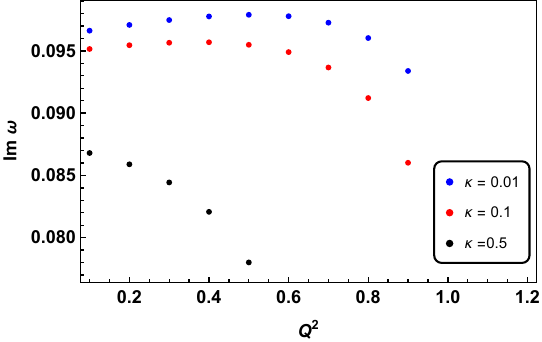}
        \caption{ Variation of Im($\omega$) with $Q^2$}
         \label{fig:neg_branch_Q_imQNM}
	\end{subfigure}
\caption{The variation of the real (left panel) and imaginary (right panel) part of the fundamental QNM with the $Q^2$ is shown for the negative branch case and $\ell=10$. The QNMs are calculated using the WKB method till $6^{th}$ order with Pad\'e approximation ($\Tilde{m} =5)$ for different values of $(\kappa,Q)$. The magnitude of Re($\omega$) increases with increasing $Q^2$ while the Im($\omega$) decreases with $Q^2$.}
\label{fig:-vechargeQNM}
\end{figure}
In Fig.(\ref{fig:neg_branch_Q_realQNM}) and (\ref{fig:neg_branch_Q_imQNM}), we show the dependence of real and imaginary components of the QNM on ($\kappa,Q^2$) for $\ell =10$, respectively. We observe that for a particular $\kappa$, as $Q^2$ increases, the frequency increases and so does the damping time. Finally, we comment on the similarity of fundamental modes for small $\kappa$ solutions with the Reissner-Nordstr\"{o}m metric as shown in Table \ref{tab:chargeQNMcompare}. We observe that indeed the negative branch black holes for small $\kappa$ have QNMs very close to the RN black hole and would require high precision observation to distinguish between them.

Table \ref{tab:QNMsummary} summarizes the dependence of the QNMs on the metric parameters for different spacetimes discussed in our work.
\begin{table}[h]
  \centering
  \begin{tabular}{|c|c|c|c|c|}
      \hline
      Branch &  $\kappa$ &  $Q^2$ & Re ($\omega$) & Im ($\omega$)\\
      \hline
      $+$ & Increasing & 0 & $\downarrow$ & $\uparrow$ \\
      $-$ & Increasing & 0 & $\uparrow$ & $\downarrow$ \\
      \hline
      $+$ & Small & Increasing & $\uparrow$ & $\uparrow$ to $\downarrow$\\
      $+$ & Large & Increasing & $\uparrow$ & $\uparrow$\\
      $-$ & Fixed & Increasing & $\uparrow$ & $\downarrow$\\
      \hline
\end{tabular}
\caption{\label{tab:QNMsummary} Table shows the dependence of the QNMs on the metric parameters for different spacetimes. $\uparrow$ and $\downarrow$ indicate respectively increasing and decreasing behaviour of the corresponding quantity with the change in the parameter. The inferences are for a fixed $\ell$.}
\end{table}

\section{Conclusion}
To summarise, gravity theories with zero metric determinant have been shown to be viable frameworks from the perspective of classical and quantum gravity \cite{Kaul:2016lhx, Kaul:2017pjw,Sengupta:2019ydf,Sengupta:2021mpf,Gera:2019ebe,Gera:2020fvo,Gera:2021dem,Gera:2021hei,Sengupta:2022ram,Sengupta:2022rbd, Sengupta:2023abg} However, the stability of the solutions in these frameworks is yet to be understood. As an initial attempt towards this quest, in this work, we study the stability and QNMs of propagating scalar fields in asymptotically flat spacetimes in the degenerate Einstein-Gauss-Bonnet gravity theory. These spacetime solutions are characterised by parameters, $\kappa$ which denotes the Gauss-Bonet coupling and the geometric charge $Q$. The metric component $g_{rr}$ or $g_{tt}$ possesses both plus and minus signs as shown in eq.(\ref{gensolution}) and the solutions can be classified into two branches corresponding to each sign. Hence, four distinct classes of spacetimes arise depending on the branch and geometric charge choices. By demanding the solutions to be black hole spacetimes, we constrain the Gauss-Bonnet coupling and geometric charge. This is achieved by demanding the presence of a horizon and any singularities in curvature scalar to be present behind the horizon. To this extent, we explicitly compute the Ricci scalar and study its singularities.  For the positive branch with zero geometric charge, the Gauss-Bonnet coupling is constrained as $0<\kappa< 8M^2$ and the Ricci scalar diverges at $r=\sqrt[3]{8M\kappa}$. A similar analysis for the negative branch without geometric charge results in $0<\kappa<M^2$. However, unlike the positive branch, the Ricci scalar diverges only at $r=0$. The above analysis is then extended to the cases with geometric charge, from which we conclude that the GB coupling in both branches restricts the range of the geometric charge. These restrictions on the parameter ranges have been discussed in detail and summarized in Table \ref{tab:spacetime}. However, unlike the zero charge case, the exact location of singularities could not be solved analytically and have been computed numerically for choices of ($\kappa,Q$), which leads to black hole solutions.

We then investigate the stability of a propagating massless scalar field in these spacetime backgrounds. The effective potentials for all cases are single barriers and hence WKB was implemented to obtain QNMs for higher $\ell$ modes. For low $\ell$ values such as  0 and 1, the stability was verified by observing the damped nature of the corresponding time domain profiles. The dependence of the QNMs on the metric parameters has been established and summarised in Table \ref{tab:QNMsummary}. From our analysis of QNMs, we observe that the modes of our black holes converge towards their GR counterparts as GB coupling tends to zero.

However, for a definite answer about the stability of these solutions, we need to check their behaviour and QNMs under gravitational perturbations, which would be a natural extension of the current work. One also wonders if these geometric charges can mimic the actual Maxwell electromagnetic charge. At least, the preliminary analysis via scalar QNMs does not rule out this possibility but requires further investigation. As for future perspectives, one can extend this analysis by adding the cosmological constant, which poses additional complexities. Alternatively, one might try to construct rotating or slow-rotating examples around these solutions and compute the shadows, QNM and other observable quantities. These results will be relevant from the standpoint of gravitational wave observation. However, this is a non-trivial endeavour and will be addressed in future.

\section{Acknowledgements}
The authors are grateful to Sayan Kar and K.G.Arun for their valuable comments on the manuscript. The authors are also thankful for the suggestions the anonymous referee gave that led to the improvement of the draft. S.G acknowledges the support of SERB project grant  CRG/2020/002035. S.G. also acknowledges the discussions and input from Kinjal Banerjee. P.D.R. acknowledges the support of a grant from the Infosys Foundation.

\appendix
\section{QNMs through Prony extraction} \label{appendix:prony}

\begin{figure}[h]
     \centering
      \includegraphics[scale=0.3]{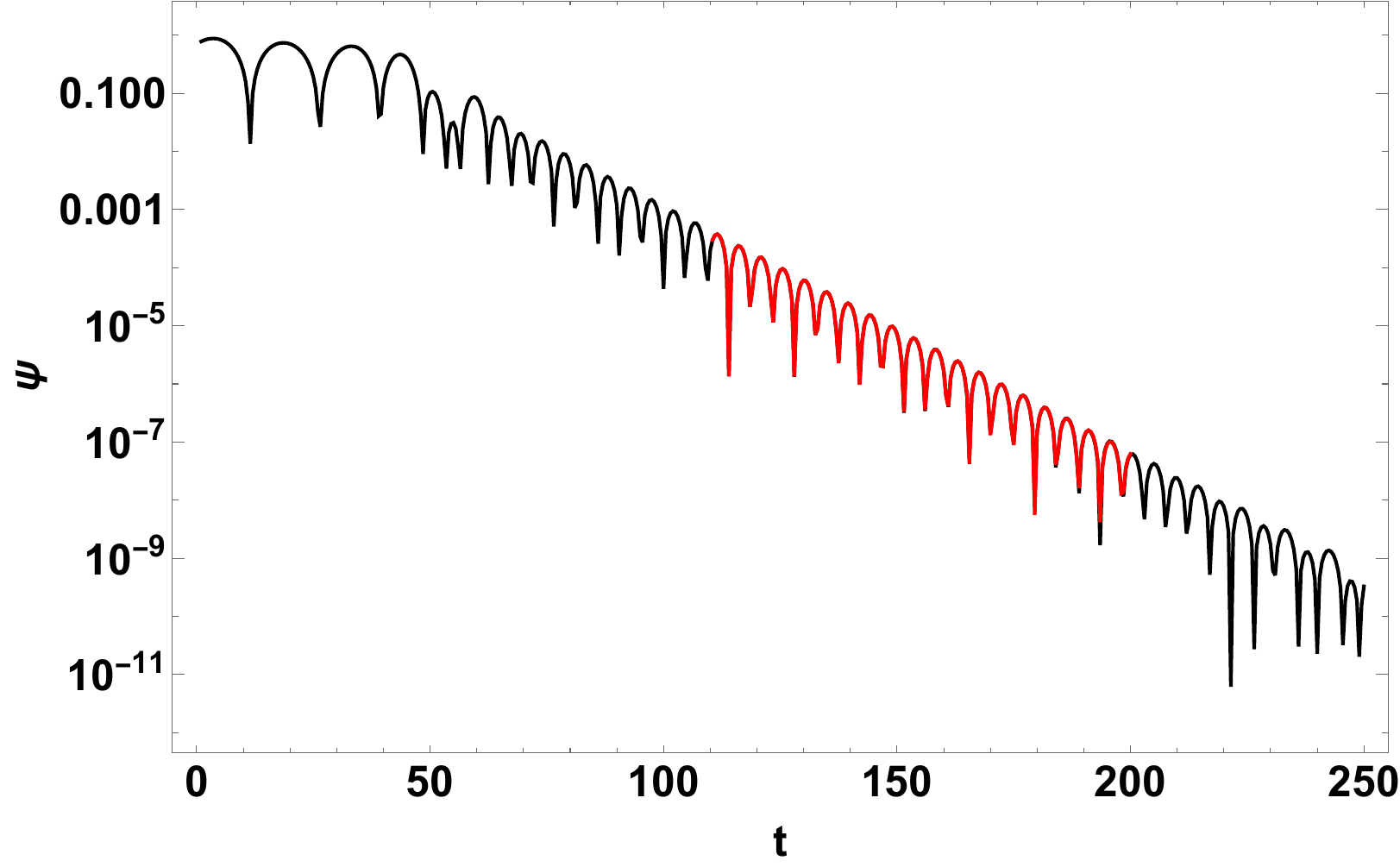}
      \caption{The time domain profile (in black) for positive branch solution with $\kappa =0.1,\, Q =0, \, \ell =3$ fitted with fundamental QNM $0.671247 -i 0.0975271$ (shown in red). The QNM is extracted from the time domain data between $t \sim 120 -200$. The QNM value is independent of slight change in the time slice and thus shows a good fit.}
      \label{fig:Prony}
\end{figure}

In this section, we verify the values of the fundamental QNMs obtained via $6^{th}$ order WKB method with Pad\`{e} approximation by performing Prony extraction for the positive and negative branch spacetimes having zero geometric charge as shown in Table \ref{tab:tab}. The Prony technique fits the time domain data with a superposition of damped sinusoidal frequencies, as discussed in \cite{konoplya_2011}. Only the most dominant fundamental mode can be recovered with good accuracy through this technique. However, computing the time domain profile for each parameter set is time consuming since the grid spacing should be small for better accuracy. Hence we apply the WKB method for QNM computation for all the spacetimes studied in our work and verify the results with Prony technique for certain sample cases. Note that while the order of the magnitude of the modes match well even for small $\ell$ values between the two techniques, the higher $\ell=3$ mode shows better matching for a specific $\kappa$. This is attributed to the fact \g{that while studying fundamental modes with overtone number $p=0$ (see eq.(\ref{eq:WKB})), WKB is less accurate for lowest $\ell$ values where $p$ and $\ell$ are comparable}. The QNM obtained from Prony extraction can also be verified by re-plotting over the time domain data, an example of which is shown in Fig.(\ref{fig:Prony}) for the positive branch case with $\ell=3,Q=0,\kappa =0.1$.
\begin{table}[h]
 \centering
 \footnotesize
   \begin{tabular}{|c|c|c|c|c|c|c|}
      \hline
      &  \multicolumn{3}{c|}{Positive branch} & \multicolumn{3}{c|}{Negative branch}\\
      \cline{2-7}
     $\ell$ & $\kappa$ & WKB &  Prony & $\kappa$ & WKB &  Prony \\ 
    \hline 
    \multirow{3}{*}{0} & 0.1 & 0.1092 -i 0.10657 & 0.0962 -i 0.10257 &  0.1 & 0.1122 -i 0.1026 & 0.1053 -i 0.09173\\
   &  6 & 0.0341 -i 0.09236 & 0.0315 -i 0.09199  & 0.9 & 0.1144 -i 0.08251 & 0.1101 -i 0.07557 \\
   \hline
    \multirow{3}{*}{1} & 0.1 & 0.2905 -i 0.09918 & 0.2899 -i 0.10034 & 0.1 & 0.2954 -i 0.09603 & 0.2933 -i 0.09873\\
   &  6 & 0.3072 -i 0.1279 & 0.2022 -i 0.1253  &  0.9 & 0.3189 -i 0.07644 & 0.3182 -i 0.0786 \\
   \hline
   \multirow{3}{*}{3} & 0.1 & 0.6704 -i 0.09792 & 0.6712 -i 0.09752 & 0.1 & 0.6805 -i 0.09497 & 0.6814 -i 0.09458\\
   & 6 & 0.5274 -i 0.13059 & 0.5255 -i 0.14253 & 0.9 & 0.7367 -i 0.07547 & 0.7378 -i 0.07513 \\
  \hline
  \end{tabular}
 \caption{\label{tab:tab} Comparison of fundamental QNM obtained from WKB method (till $6^{th}$ order with Pad\`{e} approximation) and Prony extraction for different $\kappa$ values corresponding to the positive and negative branch of $Q=0$ case. With increasing angular momentum mode $\ell$, better match is observed between the QNM values obtained by the two methods. }
\end{table}

\section{Comparison of QNMs for negative branch, $Q=0$ metric with \cite{Konoplya:2020bxa}} \label{appendix:B}
\begin{table}[h]
  \centering
  \begin{tabular}{|c|c|c|c|c|}
      \hline
   $\kappa$ &  $\alpha$ &  Values from \cite{Konoplya:2020bxa} & Our result\\ 
    \hline 
   0.05 & 0.1 & 0.595897 - i 0.188579 & 0.595897 - i 0.188578  \\ 
  0.15 & 0.3 & 0.618405 - i 0.171552 & 0.618404 - i 0.1715519\\
   0.25    & 0.5 & 0.644336 - i 0.144444 & 0.644336 - i 0.144444 \\
  \hline
\end{tabular}
\caption{\label{tab:tab2} Comparison of fundamental $\omega_{QNM}$ for different $\kappa, \ell=1, M=1/2$ values belonging to the negative branch with zero geometric charge with the values mentioned in Table I of \cite{Konoplya:2020bxa}. The QNMs quoted in column 4 are calculated using $6^{th}$ order WKB technique with Pad{\'e} approximant ($\Tilde{m} =5$).  The QNM values match upto sixth decimal place.}
\end{table}
As mentioned earlier, the metric for the negative branch with zero charge in our work matches the 4D black hole solution in EGB gravity studied in \cite{Konoplya:2020bxa} with the mapping of parameter $\alpha$ in \cite{Konoplya:2020bxa} to $2\kappa$ of our metric. Table \ref{tab:tab2} shows the values of fundamental QNM calculated for $\ell =1$ and different values of $\kappa$ or $\alpha$. Note that following \cite{Konoplya:2020bxa}, for this comparison we set $M=1/2$. The `working precision' was set to 60 while computing the QNMs in {\em Mathematica}. The values quoted in Table \ref{tab:tab2} show a good match between our results and that of \cite{Konoplya:2020bxa}. 

\bibliographystyle{JHEP}
\bibliography{ref}

\end{document}